\documentclass[a4paper]{ESASPStyle}
\usepackage{epsfig}
\begin{document}

\title{Parallel Science with Astero-seismology missions}

\author{G. Gilmore}

  \institute{Institute of Astronomy, Madingley Rd., Cambridge CB3 0HA, UK}

\maketitle 

\begin{abstract}

Eddington, COROT, MONS, KEPLER, and the other asteroseismology and
planet finding missions, obtain extremely high photometric quality
time-series data as their primary purpose.  Similar quality data are
potentially, and in some designs actually, being obtained for very
many other sources in the telescope field of view, in addition to the
primary mission targets. These parallel data, of exceptional quality
and broad scientific interest, can be made available for scientific
analysis with small system impact. This paper lists some of the most
obvious of the serendipitous research and discovery opportunities
which these parallel data will allow. The scientific potential is both
large and unique, encouraging efforts to provide these data to the
community.

\keywords{Stars: variability, binaries -- X-ray sources -- Accretion
physics -- Supernovae -- extragalactic astrophysics }
\end{abstract}

\section{Introduction}
  
High precision time-series photometry is the method of choice for
study of many astrophysical objects. These obviously include astero-
and helio-seismology, and searches for low-mass planets eclipsing
their parent stars, the subjects which have justified a series of
dedicated spacecraft projects. The generally large fields of view of
these missions, especially in their planet-finding mode of operation,
at least in principle allow time-series photometry free of the Earth's
atmospheric disturbances for many other sources in that field of view
which are not primary targets. Delivery of this parallel science data
stream for analysis is necessarily subject to the technical
constraints set by the primary mission science, especially such
factors as telemetry, sampling frequency, number of objects, on-board
processing capability, and so on.  Nonetheless, parallel science has a
proven record of discovery ({\sl pace} HST), and is manifestly a
cost-effective use of resources, so that its optimisation is a
legitimate factor in mission design.

We note here some of the scientifically most obvious, and technically
least demanding, parallel scientific studies which can produce
important and unique science from missions whose primary purpose is
asteroseismology and/or planet finding. Examples of the science
applications, and indications of the number of sources available for
study, are provided. We do not discuss here pulsating stars, wherever
they exist in the HR diagram, as such stars are core science for
asteroseismology, not additional science.

We consider in turn the following:

\begin{enumerate}
\item  Solar System Objects
\item  Young Stars
\item  Low Mass Stars
\item  Eclipsing Binaries 
\item  Accretion physics
\item  AGN and QSOs
\item  Supernovae
\item  Low surface brightness galaxies
\item Gravitational lensing
\end{enumerate}

\section{Solar System Objects}

The minor bodies of the Solar System record physical conditions in the
proto-Solar nebula, and their properties therefore shed light on the
formation of planetary systems. This is naturally a complementary
aspect of planet-discovery, a primary science goal.  Discovery and
orbital determination of near-Earth objects is also a subject of high
public interest.

Solar system objects crossing the focal plane must of course be
identified, and removed from primary target photometry. The data are
also of intrinsic interest. By detailed surveys of specific
directions, Eddington will discover substantial numbers of solar
system objects. This survey is a natural complement to larger-area
shorter integration studies. Eddington will naturally be very
sensitive to slow-moving sources, which are those hardest to detect in
normal surveys. The Eddington sensitivity limit reaches to the
brighter magnitude regions expected to contain Kuiper-belt objects,
which will have small apparent motion.

The angular motion of a typical Kuiper-belt object at $\sim90^\circ$
elongation is small: the known Kuiper-belt objects have
$d\alpha/dt=0.02-1.0$~arcsec~hr$^{-1}$ and
$d\delta/dt=0.002-1.2$~arcsec~hr$^{-1}$. That is, with the expected
spatial resolution of present mission designs, Kuiper-belt objects are
effectively stationary for times of hours, allowing long effective
integration times. The surface density of the Kuiper Belt at
$V=20$~mag is $8 \times 10^{-3}$ objects per square degree, in the
ecliptic plane, so that the very large area of sky observed is well
matched to the surface density.

\section{Young Stars}

Optical photometric variability is one of the original defining
characteristics of pre-main-sequence stars (Joy 1945). Although
pre-main sequence objects are usually identified by other, less biased
photometric and spectroscopic survey techniques now, modern
variability observations remain a valuable probe of the stellar and
circumstellar activity (e.g. Bouvier etal~1995).  Such monitoring
studies have shown that photometric variability is a diverse
phenomenon in that the observed flux can change by milli-magnitudes to
magnitudes on time scales of minutes to years, often with periodic as
well as aperiodic components. Young stellar objects are also variable
at x-ray, ultraviolet, infrared, and radio wavelengths. For periodic
stars, the variability is thought to originate mainly from cool
magnetic or hot accretion spots on the stellar surface that are
hundreds to thousands of degrees different in temperature from the
photosphere and rotate with the star.  Aperiodic variability may arise
from mechanisms such as coronal flares, irregular accretion of new
material onto the star, and temporal variations in circumstellar
extinction.

An example of the incidence of photometric variability is provided by
a recent study by Carpenter etal (2001) of a part of the Orion
Nebula Cluster.  They established the near-infrared variability
properties of pre-main-sequence stars in Orion on time scales up to 2
years.  A total of 1235 near-infrared variable stars were identified,
over 90\% of which are likely associated with the Orion~A molecular
cloud. About 30\% of their targets were detected as variable, in spite
of their relatively high detection threshold.

The variable stars exhibit a diversity of photometric behavior
with time, including cyclic fluctuations with periods up to 15 days,
aperiodic day-to-day fluctuations, eclipses, slow drifts in brightness
over one month or longer, colorless variability (within the noise
limits of the data), stars that become redder as they fade, and stars
that become bluer as they fade. The mean peak-to-peak amplitudes of
the detected photometric fluctuations were about 0.2mag in each band and 77\%
of the variable stars have color variations less than 0.05mag. 

The high surface density of such variables towards star forming
regions is illustrated in Table~1. It is worth recalling that the Sun
is presently moving through an expanding star-forming complex,
well-mapped by HIPPARCOS data (eg deZeeuw etal 1999). The local number
of young stars is higher than might be expected from simple star count
models.  Thus, while Orion~A is certainly an unusually high density
region, all low Galactic latitude directions have a large number of
young, and therefore variable, stars at apparently bright magnitudes.

\begin{table}[ht]
\caption{Variable YSO Population Associated with Orion~A}
\label{tab:cluster}
\begin{center}
 \leavevmode
    \footnotesize
    \begin{tabular}[h]{crrc}
\hline \\[-5pt]
 {Surface Density} & 
  {N}   &
  {N$_{var}$}       &
  {$f_{var}$}      \\
  {(arcmin$^{-2}$)}    & & & \\
 0.25 &  2704 & 786 & 0.29 $\pm$ 0.010 \\
 0.50 &  2148 & 627 & 0.29 $\pm$ 0.011 \\
 0.75 & 1881 & 554 & 0.29 $\pm$ 0.012 \\
 1.00 & 1488 & 445 & 0.30 $\pm$ 0.014 \\
 1.25 & 1262 & 386 & 0.31 $\pm$ 0.016 \\
 2.50 & 895 & 258 & 0.29 $\pm$ 0.018 \\
 3.75 & 752 & 202 & 0.27 $\pm$ 0.019 \\
 5.00 & 621 & 158 & 0.25 $\pm$ 0.020 \\
 7.50 & 397 &  94 & 0.24 $\pm$ 0.025 \\
10.00 & 253 &  52 & 0.21 $\pm$ 0.029 \\
      \hline \\
      \end{tabular}
  \end{center}
\end{table}

The amplitude range of the Carpenter etal variations
suggests that higher photometric precision
observations will show every young star to be intrinsically
variable. The more extreme stars have amplitudes as large as about
2mag and change in color by as much as about 1mag. The typical time
scale of the photometric fluctuations is less than a few days,
indicating that near-infrared variability results primarily from short
term processes. 

Rotational modulation of cool and hot star spots,
variable obscuration from an inner circumstellar disk, and changes in
the mass accretion rate and other physical properties in a
circumstellar disk are possible physical origins of the near-infrared
variability. Cool spots alone can explain the observed variability
characteristics in about one-half of the stars, while the properties
of the photometric fluctuations are more consistent with hot spots or
extinction changes in about one-quarter of stars.  Variations in the
disk mass accretion rate or inner disk radius, while evident, are a
minority variability source.  Comparison of the observations and the
details of variability predicted by hot spot, extinction, and
accretion disk models suggest either that another variability
mechanism may additionally be operative, or that the observed variability
represents the net results of several of these phenomena.

\subsection{ Pre-planetary systems}

An illustrative study of variability in the Pre-Main Sequence Star KH15D
induced by eclipses by circumstellar dust features has been presented
by Hamilton etal (2001). This illustrates the potential of
photometric monitoring to investigate pre-planetary structures in
stellar disks, an elegant complement to the main planet-finding
programme.

KH15D is a pre-main sequence eclipsing TTauri member of the young
cluster NGC~2264.  The orbital period is 48 days and both the length
(16d) and depth (3 mag) of the eclipse have increased with time. 
Brightening near the time of central eclipse is confirmed in 
Hamilton etal's recent data but at a much smaller amplitude than was
originally seen. 
During eclipse there is no detectable change in spectral type or
reddening, indicating that the obscuration is caused by rather large
dust grains and/or macroscopic objects. Evidently the star is eclipsed
by an extended feature in its circumstellar disk orbiting with a
semi-major axis of 0.2 AU. Continued photometric monitoring should
allow studies of the disk structure with a spatial resolution of $3
\times 10^6$ km or better.

\subsection{X-ray emission and variability}

X-ray emission has played a major role in identification of young,
nearby stars. The ROSAT All-Sky Survey (RASS) is an important tool for
discovering stellar associations and investigating their X-ray
properties. The new generation of X-ray observatories are powerful
tools for discovery and study of young stars. For example, for coronal
X-ray sources XMM has about an order of magnitude higher sensitivity
than the RASS and provides much longer exposure times allowing
continuous monitoring for more than 40 h.  Complementary and
coordinated studies between X-ray and optical monitoring is a valuable
tool to extend, and perhaps complete, the local young-star census.

\section{The lowest-mass stars}

Very low mass stars, even those which are not very young, have
recently been shown to be intrinsically variable. Spots, magnetic
activity, large convection zones, and even meteorology are all
possible explanations.  An illustrative study is that of the ultracool
dwarf BRI 0021-0214 by Martin etal (2001).

They report CCD photometric monitoring of the non-emission ultracool
dwarf BRI 0021-0214 (M9.5). Significant variability in the I-band
light curve at a period of 0.84 day is found, but appears to be
transient because it is present in the 1995 data but not in the 1996
data. They also find a possible period of 0.20 day, stable over the
year, but no periodicity close to the rotation period expected from the
spectroscopic rotational broadening ($< $0.14 day). BRI 0021-0214 is a
very inactive object, with extremely low levels of Halpha and X-ray
emission. Thus, it is unlikely that magnetically induced cool spots
can account for the photometric variability. Martin etal suggest the photometric
variability of BRI 0021-0214 could be explained by the presence of an
active meteorology that leads to inhomogeneous clouds on the
surface. The lack of photometric modulation at the expected rotational
period suggests that the pattern of surface features may be more
complicated than previously anticipated.

The magnetic Reynolds number (Rm) in the atmosphere of L dwarfs, which
describes how well the gas couples with the magnetic field, is too
small ($<<1$) to support the formation of magnetic spots.  Thus, these
authors support the idea that non-uniform condensate coverage (i.e.
clouds) is responsible for the variations.  In contrast silicate and
iron clouds form in the photospheres of L~dwarfs. Inhomogeneities in
such cloud decks can plausibly produce the observed photometric
variations.  Further evidence in support of clouds is the tendency for
variable L dwarfs to be bluer than the average L dwarf of a given
spectral type.  This color effect is expected if clear holes appear in
an otherwise uniform cloud layer.

A high level of magnetic activity in very late type stars, with
associated flaring, has also been deduced from VLA observations by
Berger (2001).

\section{Eclipsing Binary stars}

Eclipsing detached binary stars provide key tests of stellar
mass-luminosity-radius relations, tests which available stellar models
struggle to meet (eg Lebreton, Fernandes \& Lejeune, 2001; Bedin etal
2001). The corresponding stellar model tests complement the
asteroseismology science case, and in that sense are core science for
the missions, rather than additional science. We do not discuss this
further here.

It is worth noting that such studies can extend stellar structure
studies beyond single stars.  Discoveries of eclipsing detached
binaries, especially in open clusters, where independent ages can be
provided, will revolutionise those fundamental tests of stellar
evolution which include dynamical histories. As just one illustration,
the star S1082 in M67 has been recently shown to be a triple, with
inner and outer components being blue stragglers (van den Berg et al
2001). Study of such systems provides key tests of stellar evolution,
binary coalescence, and the dynamical evolution of multiple systems.

For contact binaries, the distribution of the observed light-variation
amplitudes, in addition to determining the number of
undiscovered contact binary systems falling below some photometric
detection thresholds and thus lost to statistics, serve as a tool
in determination of the mass-ratio distribution, which is very
important for understanding of the evolution and mass transfer.

Rucinski (2001) provides simulations of the expected amplitude
distribution, which show that it tends to converge to a mass-ratio
dependent constant value for sufficiently accurate photometric
data. The strong dependence of variation amplitude on mass ratio
can be used to determine the latter distribution.  Estimates based on
Baade's Window data from the OGLE project, for amplitudes a$>0.3$
mag allow determination
of the mass ratio distribution Q(q) over $0.12<q<1$, and suggest a steep
increase of Q(q) as q tends to zero.  The mass-ratio distribution can
be approximated by a power law, either $Q(q)\approx(1-q)^{a1}$ with
$a1=6\pm2$ or $Q(q)\approx q^{b1}$, with $b1=-2\pm0.5$, with a slight
preference for the former form. Both forms must be modified by the
theoretically expected cut-off caused by a tidal instability at about
$q_{min}$ =0.07-0.1. An expected maximum in Q(q), is expected to be
mapped into a local maximum in A(a) around 0.2-0.25 mag. Clearly
extension of such data below the current large photometric
detection threshold will substantially enhance such statistical
analyses.

\section{Accretion disk stellar systems}

This general class includes cataclysmic variables, dwarf novae,
high-mass and low-mass X-ray binaries, and a host of other historical
terminologies. The evolutionary state of the mass donor and of the
recipient of the accreted mass define the categories, and the
timescales and amplitudes of relevance. The basic physics is common to
all classes.

X-ray binaries (XRBs) are close binaries that contain a relatively
un-evolved donor star and a neutron star or black hole that is thought to
be accreting material through Roche-lobe overflow. Material passing
through the inner Lagrangian point moves along a ballistic trajectory
until impacting onto the outer regions of an accretion disk. This material
spirals through the disk, losing angular momentum, until it accretes onto
the central compact object, where X-rays are emitted from inner disk
regions. 

In cataclysmic variables, additional variability is categorised as a
superhump, a periodic modulation caused by a precessing eccentric
accretion disk.

X-ray novae (XNe) are mass transferring binaries in which long periods of
quiescence (when the X-ray luminosity is $\le 10^{33}$ergs s$^{-1}$) are
occasionally interrupted by luminous X-ray and optical outbursts.
XNe provide compelling evidence for the existence of stellar mass black holes,
since they can be shown to contain compact objects whose
mass exceed the maximum stable limit of a neutron star, which is $\approx
3M_{\sun}$.  Observations of the companion star in quiescence can
lead to a full understanding of the orbital parameters of the system,
including the masses of the binary components and the orbital
inclination (eg Bailyn etal 1998).

A detailed understanding of the accretion flow in these objects is of
considerable importance, since the behavior of the flow close to the
event horizon may give rise to tests of general relativity in the
strong field limits.  During their outburst cycles, XNe generally
display the complete range of spectral states, from quiescent
(``off'') to ``low/hard'' to ``high/soft'' to ``very high''.  They
therefore present unique opportunities to study all of these kinds of
accretion flows in a situation in which the geometry of the binary
system is well understood.

X--ray pulsars in High Mass X--ray Binaries (HMXBs) consist of an
accreting neutron star orbiting a (super)giant or a main--sequence
Be--type companion star.  Most known neutron stars in Supergiant XBs
emit high X--ray fluxes, driven by accretion of a roughly spherical
dense wind from the massive companion (which may be enhanced by Roche
lobe overflow). 

In contrast, the neutron stars in Be-star X-ray binaries often exhibit
transient X--ray outbursts which may occur periodically at periastron
(type I) or when the companion star undergoes a mass loss episode from
the equatorial regions due to its high rotational velocity (up to
$\sim$75\% of the break--up velocity; type II). In Be-star X-ray
binaries, the primary star is an early type star in the range 10$-$20
M$_\odot$ of luminosity class III to V, which often displays Balmer
lines, but also He\,I and metallic lines, in emission. Due to these
lines, Be stars are difficult to classify. Be X-ray binaries are young
systems, concentrated to the Galactic Plane and in the Magellanic
Clouds, complicating identification and study. Only about $\sim$20
optical counterparts have been discovered out of the $>$100 known and
10$^4$--10$^5$ expected Be/X--ray pulsars.  A illustration of the
analysis of combined optical and X-ray data for one of these sources
is provided by Reig etal (2000).

The spatial distribution of accretion disk stellar systems varies with
type. High mass B-stars are found primarily at low Galactic latitudes,
following recent star formation. Cataclysmic Variables populate the
full thin disk and thick disk volumes, while LMXBs are found
prerentially towards the Galactic bulge.

\begin{figure*}[h!t]
\begin{center}
\includegraphics[width=.8\textwidth,angle=0.]{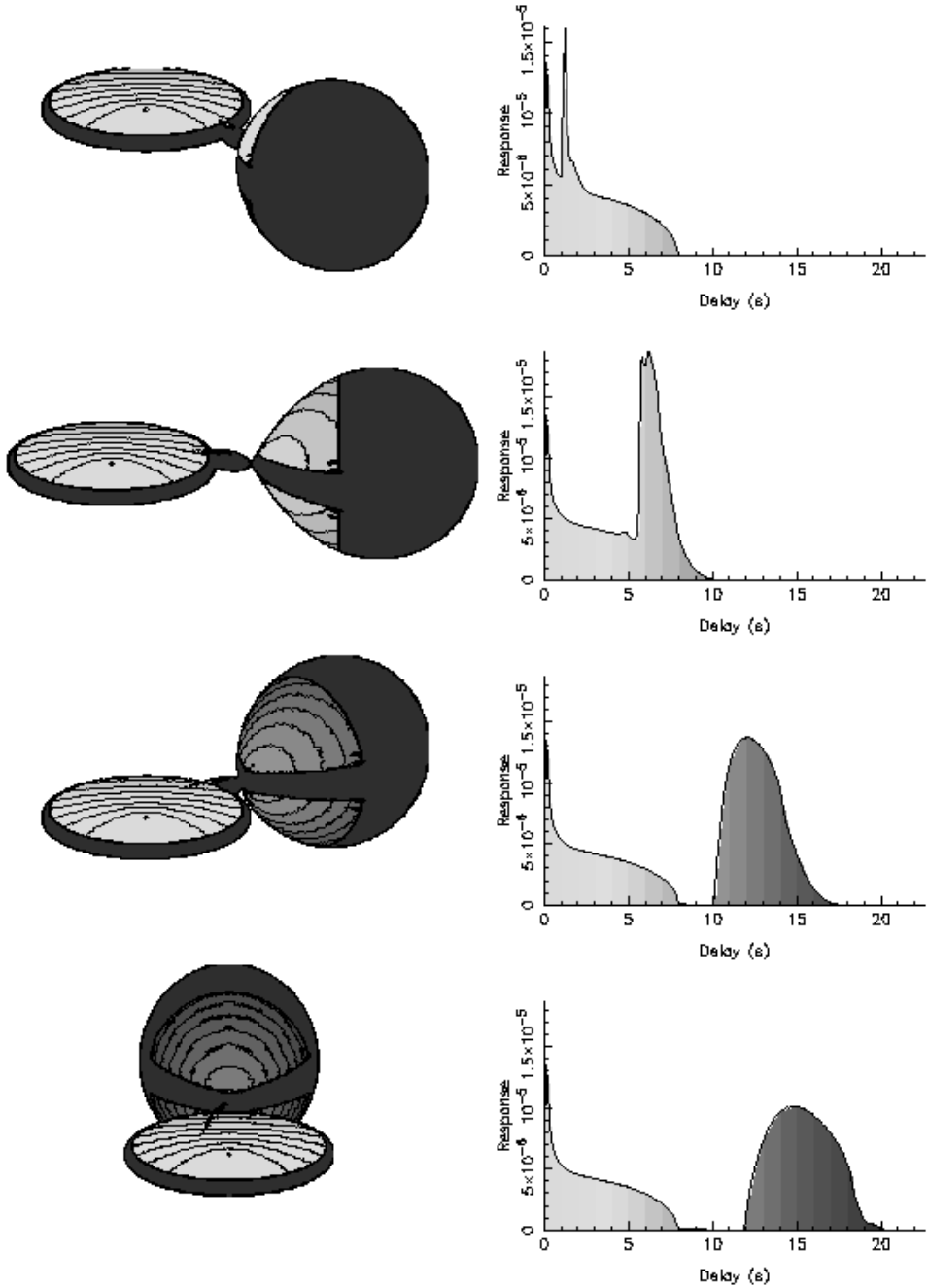}
\end{center}
\caption[]{Left, model X-ray binaries, based on the Scorpius~X-1
binary parameters, showing iso-delay surfaces projected onto the
irradiated surfaces of the binary. Right, the associated time delay
transfer functions, showing the relative contributions from the
regions highlighted in the model X-ray binaries.
The accretion disk has constant time
delays in the region 0-8 seconds, whereas the time delays from the
companion star are seen to vary sinusoidally with binary phase between
0 and 20 seconds. This figure is from O'Brien and Horne (2001).} 
\label{binaries}
\end{figure*}

In addition to discovery of probable black holes, the key physics
involved is that of accretion.
The irradiated regions of an X-ray binary are illustrated in 
Fig.~\ref{binaries}, where the left hand panels show a typical X-ray
binary, using binary parameters based on those of Nova Sco, viewed from an
inclination of 60degree (from O'Brien and Horne, 2001).

Much of the optical emission in X-Ray Binaries arises from reprocessing of
X-rays by material in regions around the central compact object. Light
travel times within the system are of order tens of seconds. Optical
variability may thus be delayed in time relative to the X-ray driving
variability by an amount characteristic of the position of the
reprocessing region in the binary, which depends in turn on the
geometry of the binary. The optical variability may be modelled as a
convolution of the X-ray variability with a time-delay transfer
function.

This time delay is the basis of an indirect imaging technique, known
as echo tomography, to probe the structure of accretion flows on
scales that cannot be imaged directly.  
The time delays observed between the X-ray and optical/UV
variability in X-ray binaries can be modelled 
to echo-map the irradiated regions, allowing study of the detailed
geometry and radiation transfer physics in the accretion disk. 

While the method of echo-tomography of X-ray binaries is still in its
infancy, current studies have shown that with just a small amount of data, from
co-ordinated observing campaigns using optical and X-ray satellite
observatories, this technique can reveal important insights into the
geometry of X-ray binaries. Furthermore this technique has the promise of
probing the structure and geometry of such systems on scales unobtainable
with any other current technique. 

An example of a recent detailed analysis of the dwarf nova/CV WG Sge,
illustrating the implications for disk structure and stability, and
for the mass transfer rate and viscosity (the `$\alpha$~-parameter'),
has been presented by Cannizzo (2001).

\subsection{Characteristic timescales}

Echo mapping has already been
developed to interpret lightcurves of Active Galactic Nuclei (AGN, see
below),
where time delays are used to resolve photoionized emission-line
regions near the compact variable source of ionizing radiation in the
nucleus. In AGN the timescale of detectable variations is days to
weeks, giving a resolution in the transfer functions of 1-10 light
days (Krolik et al 1991; Horne et al 1991). In X-Ray binaries the binary
separation is light seconds rather than light days, requiring
high-speed optical/UV and X-ray lightcurves to probe the structure of
the components of the binary in detail. The detectable X-ray and
optical variations in the lightcurves of such systems are also
suitably fast.

In the standard model of reprocessing, X-rays are emitted by material in
the deep potential well of the compact object. These photoionize and heat
the surrounding regions of gas, which later recombine and cool, producing
lower energy photons. The optical emission seen by a distant observer is
delayed in time of arrival relative to the X-rays by two mechanisms. The
first is a finite reprocessing time for the X-ray photons, 
and the second is the light travel times
between the X-ray source and the reprocessing sites within the binary
system. 

The light travel times arise from the time of flight differences for
photons that are observed directly and those that are reprocessed and
re-emitted before travelling to the observer. These delays can be up to
twice the binary separation, obtained from Kepler's third law, 
\begin{equation}
\frac{a}{c} = 9.76\mbox{s}\left(\frac{M_{\rm{x}}+M_{d}}{msolar}\right)^{\frac{1}{3}}  \left(\frac{P}{\mbox{days}} \right)^{\frac{2}{,3}}
\end{equation}
where $a$ is the binary separation, $M_{\rm{x}}$ and $M_{d}$ are the masses of
the compact object and donor star, $P$ is the orbital period. In LMXBs the
binary separation is of the order of several light seconds.

\subsection{Characteristic amplitudes}

Examples of both the considerable amplitude of variation seen, and
the very bright magnitudes which these sources reach, are provided by
V6461Sgr, and by XTE J1118+480.

V4641 Sgr = SAX J1819.3-2525 underwent a bright optical outburst on
1999 Sept. 15.7 UT, going from magnitude 14 to 8.8 in the V-band
and $Ks \simeq 13$, reaching 12.2 Crab in the X-rays.
This outburst was therefore bright, but very brief, with an e-fold
decay time of 0.6 days (Figure~2).  A radio source was resolved, making of V4641
Sgr a new microquasar (Hjellming etal 2000).
The distance of the system is probably between 3 and 8 kpc, with the companion
star being a B3-A2 main sequence star. Another possibility is that the
companion star is crossing the Hertzsprung gap (type B3-A2 IV), and in
this case the distance cited above would be the minimum distance of
the system.  The system is therefore an Intermediate or High Mass
X-ray Binary System (IMXB or HMXB).

\begin{figure}
\centerline{\psfig{file=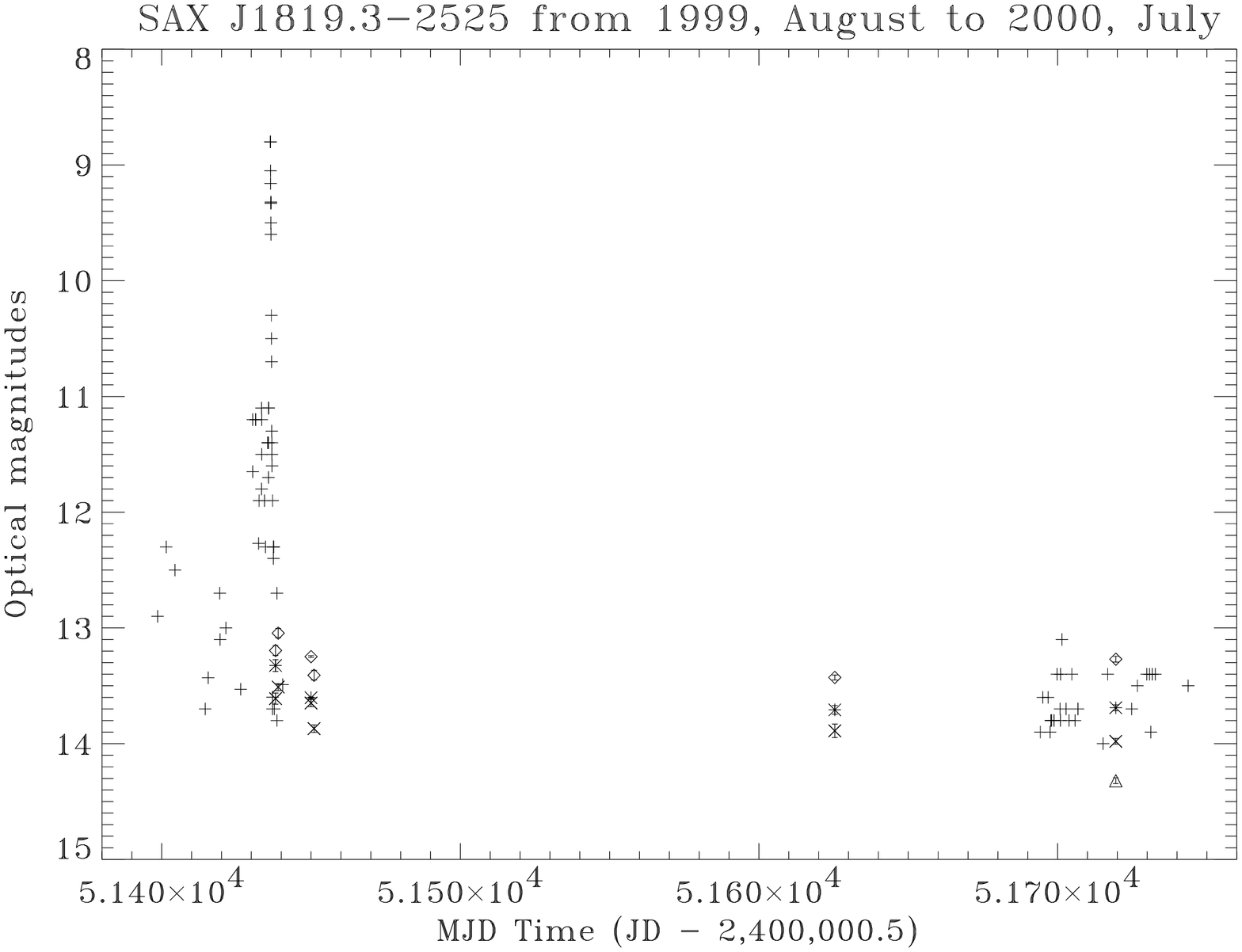,angle=+90.,width=8.9cm}}
\caption[]{Optical Observations of the X-ray transient/microquasar
V4641 Sgr . +:VSNET, $\triangle$: B, $\times$:V, $\ast$:R,
$\diamond$:I magnitudes.  The beginning of the optical activity took
place on 1999 Sept. 8 UT (= MJD 51429.5), followed by the outburst of
1999 Sept. 15.7 UT (= MJD 51437).}
%\end{figure}
%\begin{figure}[tb]
 \begin{center}
    \epsfig{file=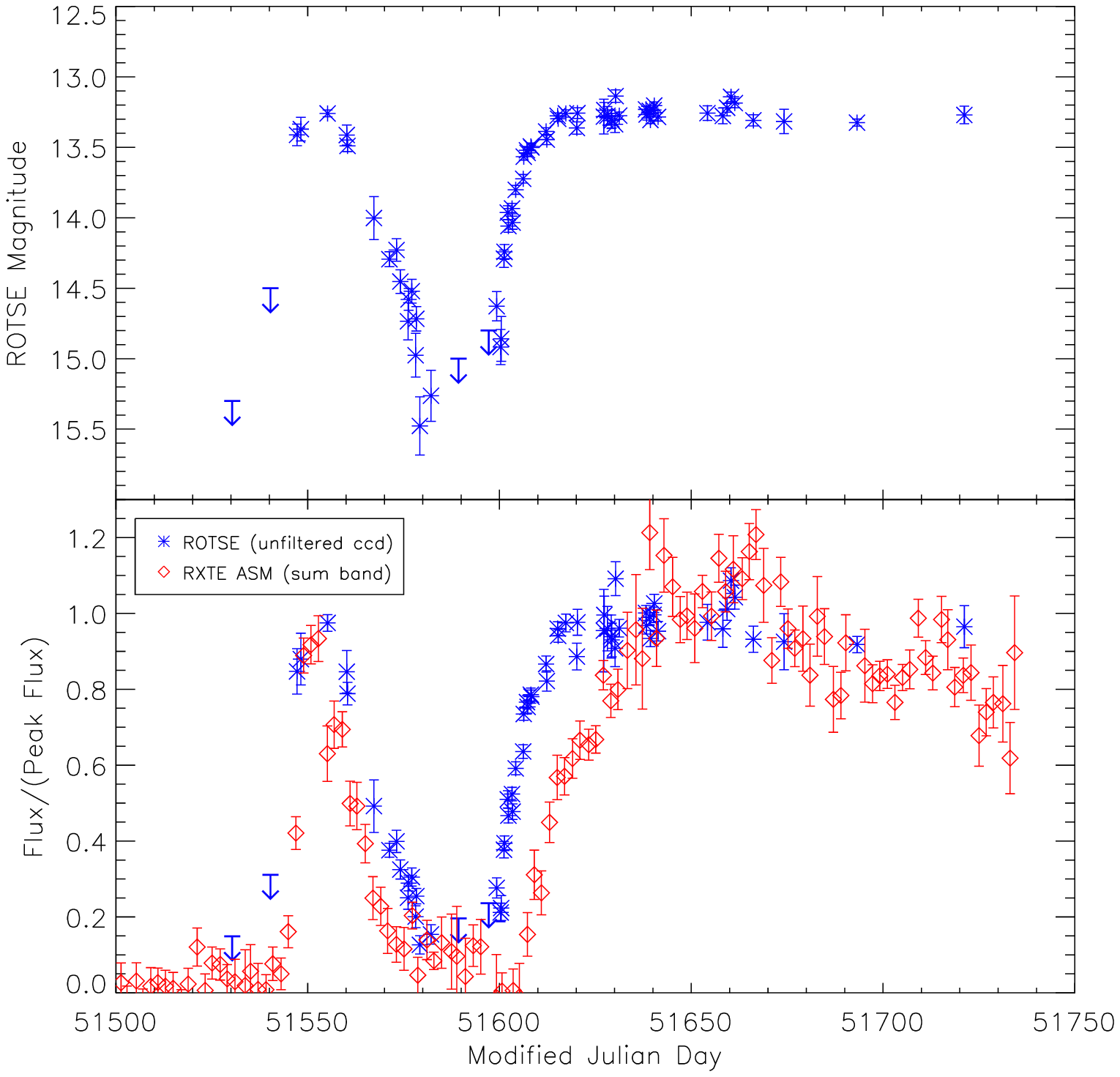, width=7cm}
  \end{center}
   \caption{The top plot shows the ROTSE-I light-curve for XTE
            J1118+480.  The lower plot is a comparison
            of the ROTSE-I and RXTE/ASM fluxes over the same time
            period.  The ASM fluxes are 2 day averages.
            The peak fluxes used for the scaling were
            $1.85\times10^{-14} erg/s/cm^{2}/\textrm{\AA}$ for ROTSE and 
            $2.8 c/s$ for the ASM. Figure from Wren etal 2001.}
\end{figure}

 The X-ray nova XTE J1118+480 exhibited two outbursts in the early
part of 2000. As detected by the Rossi X-ray Timing Explorer (RXTE),
the first outburst began in early January and the second began in
early March. Routine imaging of the northern sky by the Robotic
Optical Transient Search Experiment (ROTSE) shows the optical
counterpart to XTE J1118+480 during both outbursts. These data include
over 60 epochs from January to June 2000 (Figure~3). A search of the
ROTSE data archives reveal no previous optical outbursts of this
source in selected data between April 1998 and January 2000. While the
X-ray to optical flux ratio of XTE J1118+480 was low during both
outbursts, Wren etal (2001) suggest that they were full X-ray novae
and not mini-outbursts based on comparison with similar sources.  The
ROTSE measurements taken during the March 2000 outburst also indicate
a rapid rise in the optical flux that preceded the X-ray emission
measured by the RXTE by approximately 10 days. Using these results,
Wren etal estimate a pre-outburst accretion disk inner truncation
radius of $1.2 x 10^4$ Schwarzschild radii.

\subsection{milli-Hz Oscillations and QPOs}

X-ray pulsar binaries, such as Her X-1, can show optical mHz
quasi-periodic oscillations (QPO). In the power spectrum of Her X-1 it
appears as `peaked noise', with a coherency $\sim$2, a central
frequency of 35 mHz and a peak-to-peak amplitude of 5\%. These QPOs
are quite common, and are suggested to have a variety of causes (van
der Klis 2000), with  mHz oscillations possibly due to warping of the inner
accretion disk.  In at least some cases, QPO emission probably results
from a small hot region, possibly the inner regions of the accretion
disk, where the ballistic accretion stream impacts onto the disk

Intriguingly for the parallel science opportunities for 
a primarily asteroseismology mission, Wagoner,
Silbergleit, and Ortega-Rodriguez (2001) have introduced the analysis
concept of `Diskoseismology'. They show that one may compare
calculations of the frequencies of the fundamental g, c, and p--modes
of relativistic thin accretion disks with recent observations of high
frequency QPOs in X-ray binaries with black hole candidates. These
classes of modes encompass all adiabatic perturbations of such
disks. The frequencies of these modes depend mainly on the mass
and angular momentum of the black hole; their weak dependence on disk
luminosity is also explicitly indicated.  Identifying the recently
discovered relatively stable QPO pairs with the fundamental g and c
modes provides a determination of the mass and angular momentum of the
black hole. For GRO J1655-40, these authors derive $M=5.9\pm 1.0
M_{sun}$, $J=(0.917\pm 0.024)GM^2/c$, in agreement with spectroscopic
mass determinations.

\section{Active Galactic Nuclei}

The ultraviolet and optical continuum and the broad emission line flux
of active galaxies are known to be variable on all timescales from
hours up to years. Quantification of these variations, both their rate
of occurrence and the rate of change, provides direct study of the
inner accretion processes around massive black holes. The variations
are very broad in frequency, with much of the optical radiation being
reprocessed from higher frequencies, as in the accretion stellar
systems discussed above. Thus, perhaps the greatest impact of
Eddington observations would be in coordination with an extended
monitoring program including X-ray and $\gamma-$ray satellites, and
ground based spectroscopy. From such studies, reverberation mapping
(see the previous section) can determine  central black hole masses and
accretion rates, and the physical conditions in the accretion disk and
line emitting regions. A good review of short-term variability in AGN
in general is provided by Wagner \& Witzel (1995).

In Seyfert and lower luminosity sources, direct studies of the inner
accretion disk are feasible.  In a typical model of high-frequency
emission (Figure~4), the intrinsic emission originating in the warm
skin of the accretion disk is responsible for the spectral component
that is dominant in the softest X-ray range.  The hard X-ray line
emission requires an ionised reflecting medium, perhaps the warm
surface of the accretion disk.

\begin{figure}[h!bt]
  \begin{center}
    \epsfig{file=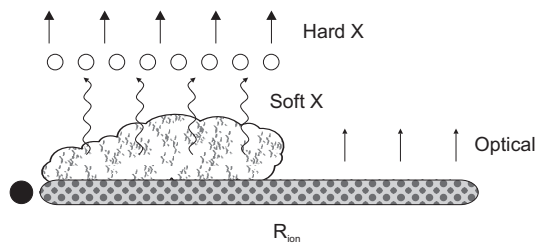, width=7cm}
  \end{center}
%\epsfxsize = 80 mm 
%\epsfbox{geometry.eps}
\caption{Geometry of the accretion flow in the Narrow line Seyfert I
galaxy PG1211+143. The optical flux is
emitted by the cold accretion disk ($T \sim 10^{4}$ K). The disk 
is the source of seed photons for the hot Comptonizing cloud 
($T \sim 10^{6}$ K, $\tau \sim 20$), which extends below the 
transition radius $R_{ion}$. The hard X--ray flux is emitted by the 
hot flare region ($T \sim 10^{9}$ K) and is partially reflected by the 
cloud ($\xi \sim 500$, $\Omega/2\pi \sim 1$). This figure is from
Janiuk, Czerny \& Madejski 2001). \label{fig:geom}}
\end{figure}

BL Lac objects, of which a well-studied example is ON~231, have the
peak of the synchrotron emission from the core source in the near
IR-optical band. Available multi-wavelength monitoring data suggest
that the occurrence of a long-term trend in the optical luminosity and
of periods of enhanced activity could be related to changes in the
innermost radio structure.  A better understanding of these phenomenon
requires both optical monitoring and VLBI mapping.

\begin{figure}[h!t]
  \begin{center}
    \epsfig{file=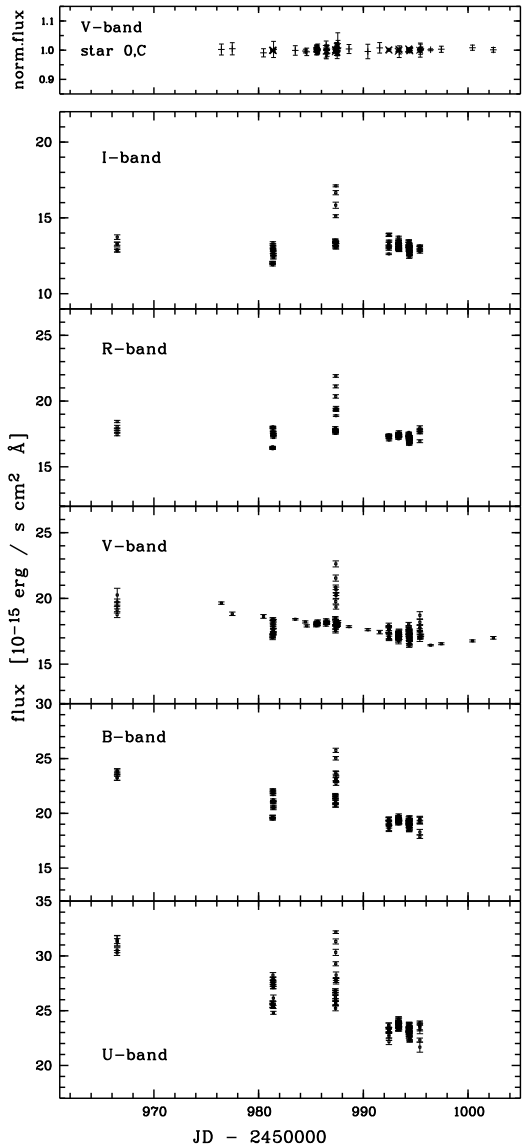, width=7cm}
  \end{center}
%
%\resizebox{\hsize}{!}{\includegraphics[bb=33 35 373 780]{MS10482f1.eps}}
%{\epsfig{figure=MS10482f1.eps,width=8.0cm,clip=t}}
\caption{Optical broad-band light curves for NGC\,5548. Fluxes 
in broad-band $U$,\,$B$,\,$V$,\,$R$, and $I$  are in units of 
$10^{-15}$\,erg s$^{-1}$\,cm$^{-2}$\,\AA$^{-1}$.
In the top panel the normalized V-band flux of star 0 (+) and 
C (x) is shown which was contant within $\sim$0.57\,\% .(From Dietrich
etal, 2001)}
\end{figure}

\begin{figure}[h!t]
  \begin{center}
    \epsfig{file=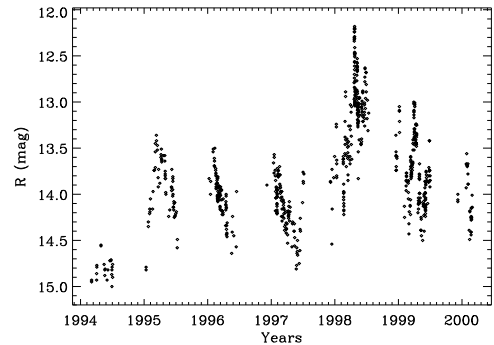, width=7cm}
  \end{center}
\caption{The light curve of ON~231 in the R (Cousins)
band after 1994, from Massaro etal 2001.)}
\end{figure}

Variations on timescales of hours and even less have been observed in
AGN.  To study the physical processes which are responsible for the
observed spectral energy distribution of an active galactic nuclei
(AGN) multiwavelength monitoring campaigns have proven to be an
excellent tool, (cf. Alloin etal 1994 for an AGN watch overview and
Gilmore (1980) for an earlier example).  Thus, over the last decade,
several large space-based and ground-based monitoring programs have
been undertaken for nearby AGN.

Historic light curves of some bright BL Lac objects have shown that
fast luminosity fluctuations (typical of this class of AGNs) are
frequently superimposed on long-term trends of relatively large
amplitude.  The origin of these long-term changes is not fully
understood.  A possibility is that the jet, pointing very close to the
observing direction, undergoes strong instabilities and oscillations:
if relativistic plasma blobs are ejected from the central source with
a small diversity of orientations, substantial ranges in relativistic
boosting factors will naturally occur. 
Another possible scenario is that of a slowly precessing jet
approaching the observer's line of sight over the past few decades.
The progressive increase of the beaming factor would then be
responsible for the mean brightening or fading optical trends shown by 
many BL Lac sources.

 Correlation studies of optical outbursts with higher frequency data
and radio VLBI maps are required to discern the true evolution of
these structures.

Extremely rapid changes of large amplitude occur in AGN
luminosity. Many examples exist.  Data for June 1998 on the Seyfert\,1
galaxy NGC\,5548, continuously monitored in the optical since late
1988 by the international AGN watch consortium are shown in figure~5
(from Dietrich etal 2001).  The broad-band fluxes (U,B,V), and the
spectrophotometric optical continuum flux monotonically decreased in
flux while the broad-band R and I fluxes and the integrated
emission-line fluxes of Halpha and Hbeta remained constant to within
5\%. On June 22, a short continuum flare was detected in the broad
band fluxes. It had an amplitude of about 20\% and it lasted only 90
min. The broad band fluxes and the optical continuum appear to vary
simultaneously with the EUV variations.

The extreme variability of the much-studied BL Lac object ON231 has
been illustrated recently by Massaro et al (2001). They have analysed
radio images of ON231 (W Com, 1219+285), showing remarkable new
features in the source structure compared to those previously
published. The R band luminosity evolution in the period 1994--1999 is
shown in figure~6.  These authors identified the source core in their
VLBI radio images with the brightest component having the flattest
spectrum. A consequence of this assumption is the existence of
two--sided emission in ON231 not detected in previous VLBI images. A
further new feature is a large bend in the jet at about 10 mas from
the core. The emission extends for about 20 mas after the bend, which
might be due to strong interaction with the environment surrounding
the nucleus. They suggest interpretations relating the changes in the
source structure with the optical and radio flux density variation in
the frame of the unification model.

\subsection{AGN: are any visible?}

Since AGN are primarily studied at high Galactic latitude, it is
necessary to ask if any at all will be visible to a stellar/planetary
mission like Eddington. While this is quantifiedd in the next section,
an example of what is viable is provided by the recent identification
of 3EG J2016+3657, an EGRET blazar behind the Galactic Plane.

Halpern etal (2001) recently identified the blazar-like radio source
G74.87+1.22 (B2013+370) as the counterpart of a high-energy
$\gamma$-ray source in the Galactic plane.  However, since most blazar
identifications of EGRET sources are only probabilistic in quality
even at high Galactic latitude, and since there also exists a
population of unidentified Galactic EGRET sources, they obtained
additional evidence to support identification of this source as a
blazar.  Their new observations provide a complete set of
classifications for the 14 brightest ROSAT X-ray sources in the
relevant error circle (Figure~7), of which B2013+370 remains the most likely
source of the $\gamma$-rays.  They also obtained further optical
photometry of B2013+370 itself which shows that it is variable,
providing additional evidence of its blazar nature.

\begin{figure}[h!t]
  \begin{center}
    \epsfig{file=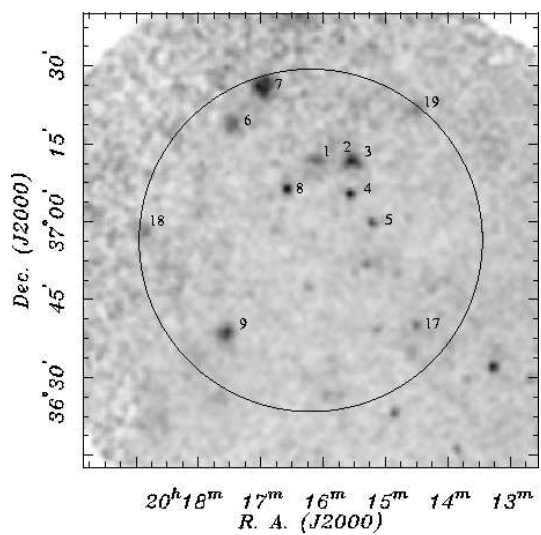, width=7cm}
  \end{center}
\caption{ROSAT PSPC image and 95\% confidence error circle of the
source G74.87+1.22, taken from Mukherjee et al. (2000.  The X-ray
sources are numbered as in that paper.  All sources are point-like
except for \#1, which is the SNR CTB 87.\label{fig7}}
\end{figure}

Interestingly,
this field contains, in addition to the blazar, the plerionic
supernova remnant CTB 87, which is too distant from the field centre
to be the EGRET source, and three newly discovered cataclysmic
variables, all five of these X-ray sources falling within
$16^{\prime}$ of each other.  This illustrates the very large surface
density of astrophysically interesting high-energy sources in the
Galactic plane.

\section{Supernovae and Gamma-ray bursts}

Supernovae are of increasing importance as cosmological probes, as
well as continuing of interest as tracers of recent (Type~II) and past
(Type~I) star formation histories, and as the origin of much of the
chemical elements. In addition they are of considerable intrinsic
interest, as efforts continue to understand the diversity of light
curves and rates of different supernova types.  At present, there is
no reliable {\sl ab initio} model of a supernova explosion: rather,
`artificial' explosions are introduced into the models, with a radial
structure adjusted {\sl ad hoc} to reproduce observed chemical element
ratios.  Supernova progenitor models remain in need of more direct
observational constraints (eg Smartt etal 2001).

\begin{figure}[ht]
  \begin{center}
    \epsfig{file=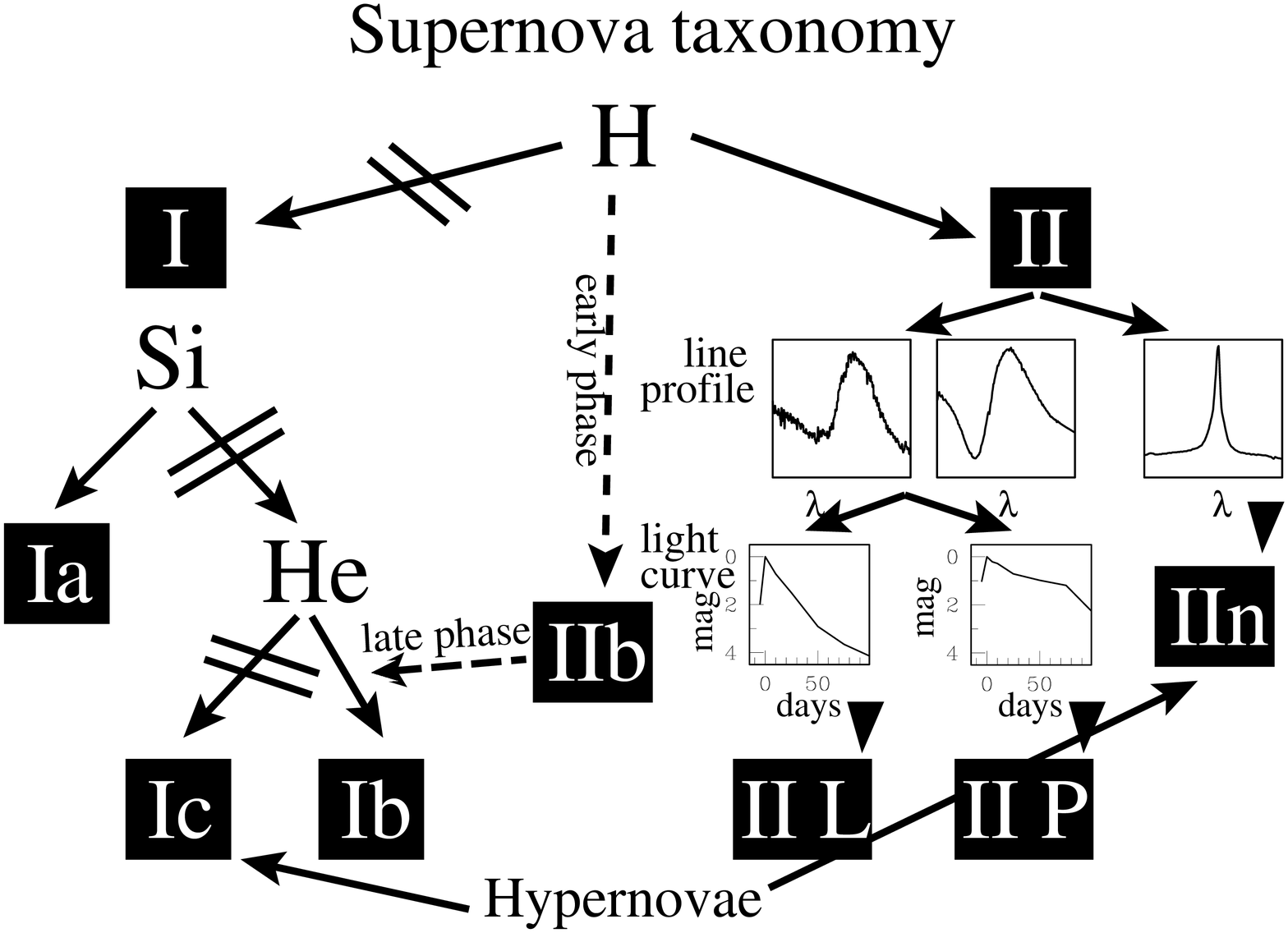, width=7cm}
  \end{center}
%\resizebox{\hsize}{!}{\includegraphics*{/SNae/taxonomy.eps}}
\caption{Classification of SNe requires not only 
identification of  specific features in the early spectra, but also analysis
of line profiles, luminosity and spectral evolution (from Cappellaro
and Turatto 2000)}
%\end{figure}
%\begin{figure}[ht]
  \begin{center}
    \epsfig{file=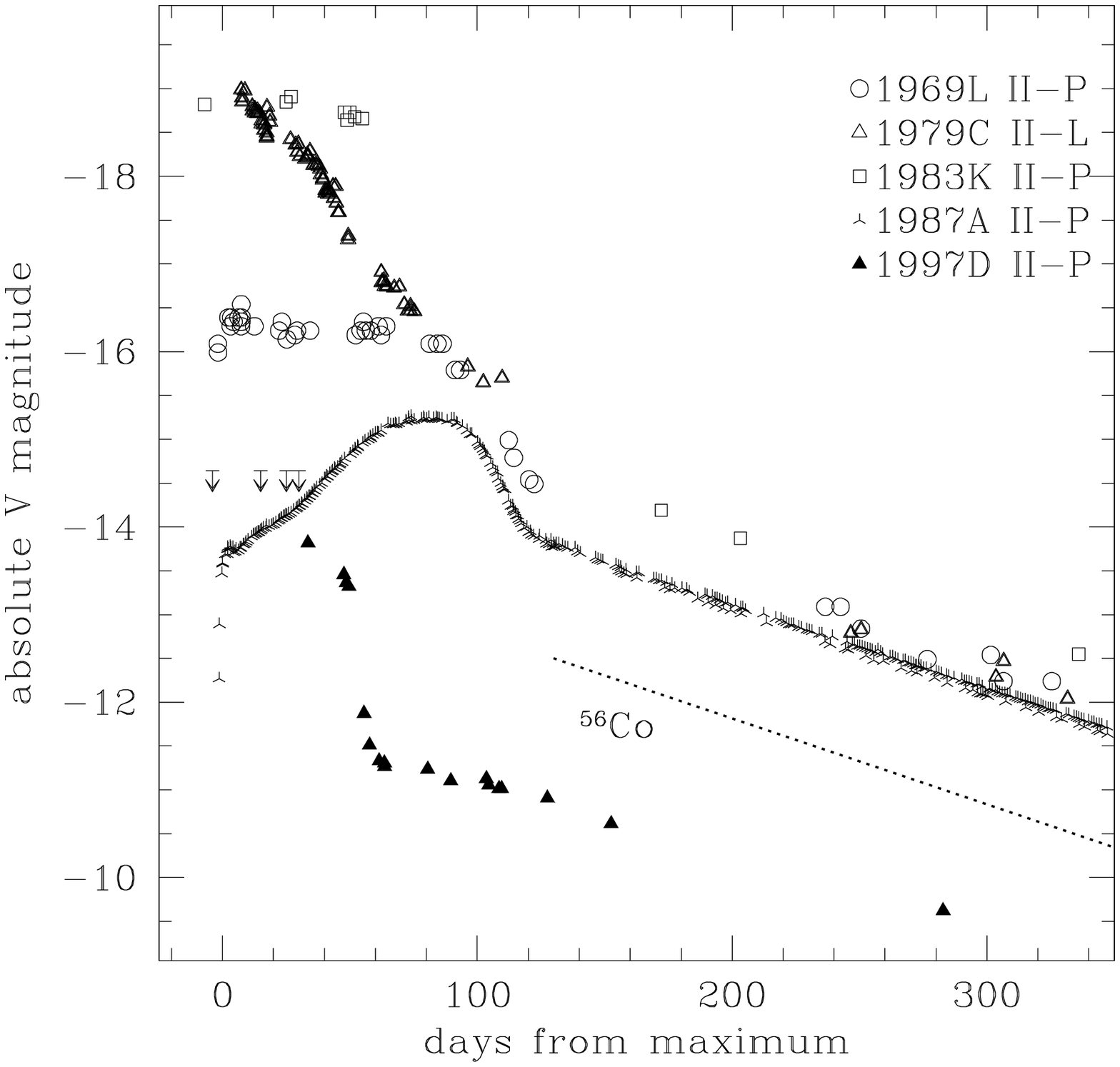, width=7cm}
  \end{center}
%
%\resizebox{11cm}{!}{\includegraphics*{/SNae/ii_lc.eps}}
\caption{Representative light curves of SN~II. The dotted line is the 
expected luminosity decline rate when the light curve is powered  
by the decay of $^{56}$Co. Note the near-complete absence of data at
early times (from Cappellaro
and Turatto 2000).}\label{iilc}
\end{figure}

One of the key ways in which Eddington will contribute here,
and in which no other observational program will be competitive with
Eddington results, is in definition of the early-time light-curves. It
is the structure of the very early light curve which provides clues to
the symmetry and (hopefully) mechanism of explosion, crucial
information at present completely lacking.  Eddington will
provide time-resolved photometry during the few hours to few days in
which different supernovae brighten, allowing direct tests of the
structure of the explosive shocks with observations. 

We emphasise that no rapid or special analysis of the data is
required, so that there is system impact only on telemetry: supernovae
will be detected in the focal plane during normal observations, and
can be recovered after the event, provided data are kept.

Two further points are worthy of note. That subset of supernovae,
often called hypernovae, which cause gamma-ray bursts, is of especial
interest. It remains unknown just what are the statistics of these
sources, and what if any intermediate types of these sources
exist. Eddington has a less than twenty-percent chance of discovering, and
providing full early light-curve data for, such an event, assuming
observations in fields where total line of sight extinction is less
than a few magnitudes. This does
however imply a system impact: the (few known) optical counterparts of 
gamma-ray bursts are very short-lived, so that the whole focal plane
needs to be analysed for sources each time it is read. 

\subsection{Supernova rates}

Present supernova rates are poorly known, while the detectivity will
depend on both the Eddington observational mode and the distance from
any bright stars. 

A recent compilation of relevant rate data has been
provided by Cappellaro \& Turatto (2000). These data (Table~2) show that one
supernova occurs per $10^{10}{\rm L}_{\odot,{\rm B}}/ 100 {\rm yr}$.
Thus, an estimate of performance is
possible. This unit luminosity is comparable to that of a typical
galaxy observed in a magnitude-limited sample. A `typical' supernova
is near maximum brightness for about one-month. Thus, one supernova is
visible in any month from monitoring of about 1000 galaxies;
given a combination of collecting area and field of view which
includes 1000 galaxies, one supernova per month will be found.
Scaling from extant statistics, Eddington will discover,
and provide a complete fully-sampled early light curve for, one
supernova for every 30 days of operations, for operation away from the
highest extinction lines of sight. We emphasise that no such light
curve exists, or is realistically obtainable, at present.

\begin{table}[ht]
\caption{Local Supernova rates. Units are $SNu = {\rm SN}/10^{10} 
{\rm L}_{\odot,{\rm B}}/ 100 {\rm yr}$. h=H$_0$/100.  }\label{rate}
  \begin{center}
    \leavevmode
    \footnotesize
 \begin{tabular}[h]{ll}
 \hline \\[-5pt]
galaxy    & {SN rate}\\
type      & (all types) \\
 \hline \\[-5pt]

E-S0      &   $0.32\pm.11~h^{2}$   \\
S0a-Sb    &   $1.28\pm.37~h^{2}$\\
Sbc-Sd    &   $2.15\pm.66~h^{2}$ \\
          &   \\
All types       &   $1.21\pm.36~h^{2}$\\
 \hline \\[-5pt]
\end{tabular}
\end{center}
\end{table}

\subsection{QSO and Galaxy counts}

A key question of course, for supernovae and all extragalactic parallel
science, is whether any sources will be visible. Table~3
presents summary QSO and galaxy counts, showing that considerable numbers of
galaxies will be in the field of view, except when the primary science
field has very high extinction. The QSO counts are taken from Hartwick
\& Schade (1980), the galaxy counts from Gardner et al (1997).
In using this table, it is important to note that the supernovae are
(in general) much brighter than the galaxy.

\begin{table}[ht]
\caption{Galaxy counts, QSO counts: units are log N/mag/sq deg}
  \begin{center}
    \leavevmode
    \footnotesize
 \begin{tabular}[h]{lllc}
 \hline \\[-5pt]
mag    & galaxies & galaxies & QSOs \\
 & B-band & I-band & B-band \\
 \hline \\[-5pt]
15.25 & 0.15 & 1.17 & -2.33 \\
15.75 & 0.15 & 1.35 & -1.93 \\
16.25 & 0.67 & 1.67 & -1.70 \\
16.75 & 0.79 & 1.97 & -1.15 \\
17.25 & 1.12 & 2.18 & -0.64 \\
17.75 & 1.44 & 2.45 & -0.13 \\
18.25 & 1.64 &      &  0.13 \\
18.75 & 1.88 &      &  0.77 \\
19.25 & 2.08 &      &  0.95 \\
19.75 & 2.32 &      &  1.20 \\
 \hline \\[-5pt]
\end{tabular}
\end{center}
\end{table}

\section{Low surface brightness galaxies}

Low surface brightness galaxies (LSB) are of considerable intrinsic interest
as cosmological and galaxy formation probes. Their existence alone is
enough of a puzzle, and they may make up  a significant part of all
the stars in galaxies. LSB galaxies are hard to detect against the sky
brightness. In principle, space observations, where the sky is very
much darker, are more sensitive, while the wide field relatively large
sky area per pixel of current satellite designs is ideal to detect
very low surface brightness sources. However, on current designs, data
from Eddington and the related missions is read-noise limited on sky
areas. Thus, obtaining very deep observations by stacking data is not possible.
But indirect LSB detection, and more generally, detection of any
stellar population which does not follow the distribution of high
surface brightness galaxies, is certainly possible.

A bonus of the Eddington parallel science is that the fields are
chosen for other reasons: thus Eddington will be sensitive to
supernovae in putative Low Surface Brightness galaxies, as well as to
field supernovae, if the descendants of the earliest stars
(Population~Zero) are indeed, as expected in some models, distributed
in space more like dark matter than luminous galaxies.  Any such
discovery, or tight limits, would be valuable.  Knowledge of the local
mass distribution in the Universe has implications for the peculiar
velocity field, the direction and amplitude of the Local Group
acceleration, the determination of parameters such as $\Omega_0$ and
$H_0$, and on the understanding of the formation and evolution of
groups of galaxies (e.g., Peebles 1994).

\begin{figure}[ht]
  \begin{center}
    \epsfig{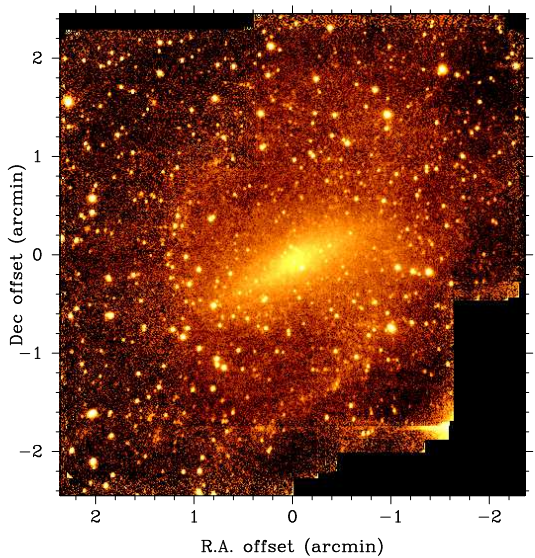}
  \end{center}
\caption{ $H$-band mosaic of Dw1 displayed on a logarithmic 
scale. The orientation is north up, east to the left. The field
of view is $4.7\arcmin \times 4.9\arcmin$. (from Ivanov etal 1999)}
%\plotfiddle{/EX/Dw1.ps}{425pt}{0}{80}{80}{-270}{-80}
\end{figure}

Is it possible to find such systems, when observations are primarily
at low Galactic latitude? Indeed yes, LSB galaxies are already known at
very low Galactic latitude.

Dwingeloo 1 (Dw1) is a large SBb/c galaxy, discovered both in a
systematic H\,{\sc i} emission survey of the northern part of the
Milky Way in search of obscured galaxies in the Zone of Avoidance by
Kraan-Korteweg et al. (1994), and independently by Huchtmeier et
al. (1995).  Interestingly, Dwingeloo 1 is known to contain a bright
X-ray source (Reynolds et al. 1997), arguably a SuperEddington Source
(sic), as luminous as a 10 M$_{\odot}$ black hole accreting at close
to its Eddington limit; however, the nature of these super-Eddington
sources remains unclear.  An optical counterpart to the
super-Eddington X-ray source, NGC 5204 X-1 has been recently
suggested, confirming the need for continuing optical studies of these
sources and these galaxies.

\section{Gravitational Lensing}

\subsection{Microlensing}

Gravitational microlensing of background stars in the Galactic bulge
by foreground stars in the Galactic disk is a rare
phenomenon. Nonetheless, it happens, and is observed. The
characteristic microlensing signals are the light curve, lack of
colour variability, and photometric stability outside the lensing
event. This requires photometry over longer timescales than are
appropriate for the present proposed missions, and multi-colour data. For planet
finding missions, microlensing will be a noise source rather than a signal.

One possibly interesting application where planet searching modes can
study microlensing directly, is in pixel microlensing. This method
considers fluctuations in the integrated light from all the unresolved
stars in a pixel. A critical requirement is point spread function
(psf) and pointing stability, maximising comparison of like with like
data.  If there is an unresolved background galaxy in any field, pixel
microlensing will be a sensitive opportunity, given the psf stability,
which vastly exceeds that available from the ground, so long as the
spacecraft jitter is small (cf Calchi Novati etal 2001).

\subsection{Macrolensing}

Gravitational macro-lensed systems, such as the famous `Einstein Cross',
are known and are variable. Analysis of the variability delays between
the image components can be of profound significance for cosmological
studies. However, the complexity of the image structure, together
with the natural several-year timescales, and the rareness of such
systems, precludes this as a major
contribution of the present missions.

\section{Conclusions}

Asteroseismology and planet transit searching necessarily produce
high-quality time-series photometry for the primary target
stars. Where system considerations allow, the equally high-quality
photometry for other objects in the field of view can provide
excellent science in many fields. Some of the most obvious are noted here.

While direct star-count studies of Galactic structure are not an ideal parallel
science priority, we also recall however the recent substantial advances
in mapping phase-space substructures in the outer Galactic halo which
followed identification of a large sample of faint RRLyrae stars (eg
Vivas etal, 2001). The significance of pulsating variables
as distance indicators, RR Lyraes for old stellar populations,
Cepheids for young populations, and the relevance of those distances
for studies of galactic structure, is worth consideration.


\begin{thebibliography}{}
%\addcontentsline{toc}{section}{References}

\bibitem[]{Alloin94} 
Alloin, D., Clavel, J., Peterson, B.M., Reichert, G.A., \& Stirpe, G.M.,
1994, in Frontiers of Space and Ground-Based Astronomy, ed.\ 
W.\,Wamsteker, M.S.\,Longair, \& Y.\,Kondo
(Dordrecht: Kluwer), p.\ 423

\bibitem[]{} Bailyn, C., Jain, R., Coppi, P. \&  Orosz, J. 1998 ApJ
499 367

\bibitem[]{} Bedin, L.R., Anderson, J., King, I.R., \&
Piotto, G., 2001 {ApJL} {560} L75

\bibitem[]{} Berger, E., 2001  astro-ph/0111317

\bibitem[]{} Bouvier, J., Covino, E., Kovo, O., Martin, E. L.
 Matthews, J. M., Terranegra, L., Beck, S. C.  1995 {A\&A} {299} 89

\bibitem[]{} Calchi Novati, S., etal 2001  astro-ph/0110706

\bibitem[]{} Cannizzo, J.K. 2001 ApJ 561 L175

\bibitem[]{} Cappellaro, E., \& Turatto, M., 2000  astro-ph/0012455

\bibitem[]{}Carpenter, J. M., Hillenbrand, L. A.,  Skrutskie,
M. F. 2001 {AJ}  {121} 3160

\bibitem[]{} Dietrich, M.,  etal 2001 A\&A 371 79

\bibitem[]{} Gardner, J.P., Sharples, R.M.,
Carrasco, B.E., \& Frenk, C.S.,  1997 {ApJ}  480 L99

\bibitem[]{} Gilmore, G., 1980 MNRAS 190 649

\bibitem[]{} Hamilton, C.M., Herbst,W.,  Shih, C., and 
Ferro, A.J.,  2001{ApJ} {554} 201

\bibitem[]{} Halpern, J. P., Eracleous, M., Mukherjee, R.. \&
 Gotthelf, E. V. 2001   ApJ 551 1016

\bibitem[]{} Hartwick, D \& Schade, D., 1980 {ARAA} {28} 437

\bibitem[]{} Hjellming,R.,  etal 2000 ApJ 544 977

\bibitem[]{kdh91} {Horne}, K., {Welsh}, W.,  and {Peterson}, B.M.
1991, ApJL, 367, L5 

\bibitem[]{} Huchtmeier, W.K., Lercher, G., Seeberger, R., 
Saurer, W. \& Weinberger, R. 1995, A\&A, 293, L33

\bibitem[]{} Ivanov, V.D., Alonso-Herrero, A.,
 Rieke, M.J., \& McCarthy, D, 1999 AJ 118 826

\bibitem[]{} Janiuk, A., Czerny, B., Madejski, G. 2001 ApJ 557 408

\bibitem[]{} Joy, A.H. 1945 {ApJ} {102} 168

\bibitem[]{} Kraan-Korteweg, R.C., Loan A.J., Burton, W.B. 
et al. 1994, Nature, 372, 77

\bibitem[]{krolik91} J. H. {Krolik}, Keith {Horne}, T. R. {Kallman},
M. A. {Malkan}, R. A. {Edelson} and G. A. {Kriss}, 1991, ApJ, 371, 541 

\bibitem[]{} Lebreton, Y., Fernandes, J., Lejeune, T. 2001 {A\&A} {374} 540 

\bibitem[]{} Martin, E.L., Zapatero Osorio, M.R., and Lehto,
H.J., 2001 {ApJ} {557} 822

\bibitem[]{}  Massaro, E., Mantovani, F., Fanti, R., Nesci, R.,
 Tosti, G., \& Venturi, T. 2001 A\&A 374 435

\bibitem[]{} Mukherjee, R., Gotthelf, E. V., Halpern, J., Tavani,
M. 2000 ApJ 542 740

\bibitem[]{} O'Brien, K., \& Horne, K., 2001 astro-ph/0104428 

\bibitem[]{} Peebles, P.J.E. 1994 Principles of Physical Cosmology
(Princeton Univ Press).

\bibitem[]{} Reig, P., Negueruela, I., Coe, M. J., Fabregat, J.,
 Tarasov, A. E., \&  Zamanov, R. K. 2000 {MNRAS} 317 205

\bibitem[]{} Reynolds, C. S. Loan, A. J. Fabian, A. C.
 Makishima, K. Brandt, W. N. \& Mizuno, T. 1997 MNRAS 286 349.

\bibitem[]{}  Rucinski, S., 2001 {AJ} {122} 1007


\bibitem[]{} Smartt, S.J., Gilmore, G.F., Trentham, Neil,
 Tout, C.A., \&  Frayn, C.M.,  2001 ApJ 556 L29

\bibitem[]{} van den Berg, M., Orosz, J., Verbunt, F., \& Stassun, K.,
2001 {A\&A} {375} 375

\bibitem[]{} van der Klis, M., 2000 {ARAA} 38 717

\bibitem[]{} Vivas, A.K.,  R. Zinn, P. Andrews, C. Bailyn, C. Baltay,
P. Coppi, N. 
  Ellman, T. Girard, D. Rabinowitz, B. Schaefer, J. Shin, J. Snyder, S. Sofia,
  W. van Altena, C. Abad, A. Bongiovanni, C. Briceno, G. Bruzual, F. Della
  Prugna, D. Herrera, G. Magris, J. Mateu, R. Pacheco, Ge. Sanchez, Gu.
  Sanchez, H. Schenner, J. Stock, B. Vicente, K. Vieira, I. Ferrin, J.
  Hernandez, M. Gebhard, R. Honeycutt, S. Muffson, J. Musser, A. Rengstorf (the
  QUEST Collaboration) 2001 ApJ 554 L33

\bibitem[]{} Wagoner, R.V., Silbergleit, A.S. and 
Ortega-Rodriguez, M. 2001 ApJ 559 L25

\bibitem[]{} Wagner, S.J., \& Witzel, A. 1995 {ARAA} {33} 163

\bibitem[]{} Wren, J., etal  2001 ApJ 557 L97

\bibitem[]{} de Zeeuw, P. T., Hoogerwerf, R., de Bruijne, J. H. J.,
 Brown, A. G. A., \& Blaauw, A., 1999 AJ 117 354

\end{thebibliography}
\end{document}